\documentclass{article}
\usepackage{amsmath}
\usepackage{braket}
\usepackage{tikz}
\usepackage{authblk}
\usepackage{caption}
\usepackage{subcaption}
\usepackage{lineno}
\usepackage{hyperref}
\usepackage{graphicx} 
\usepackage[backend=biber,style=alphabetic,sorting=ynt]{biblatex}
\title{Momentum exchange method for quantum Boltzmann methods}
\author{Merel A. Schalkers\footnote{Corresponding author: \url{m.a.schalkers@tudelft.nl}} \and
        Matthias Möller}
\affil{Department of Applied Mathematics,\\ Delft University of Technology, The Netherlands}
\date{April 2024}

\begin{document}

\maketitle

\begin{abstract}
The past years have seen a surge in quantum algorithms for computational fluid dynamics (CFD). These algorithms have in common that whilst promising a speed-up in the performance of the algorithm, no specific method of measurement has been suggested. This means that while the algorithms presented in the literature may be promising methods for creating the quantum state that represents the final flow field, an efficient measurement strategy is not available. 
This paper marks the first quantum method proposed to efficiently calculate quantities of interest (QoIs) from a state vector representing the flow field. In particular, we propose a method to calculate the force acting on an object immersed in the fluid using a quantum version of the momentum exchange method (MEM) that is commonly used in lattice Boltzmann methods to determine the drag and lift coefficients. In order to achieve this we furthermore give a scheme that implements bounce back boundary conditions on a quantum computer, as those are the boundary conditions the momentum exchange method is designed for.
\end{abstract}

\section{Introduction}
Computational fluid dynamics is one of the most frequently applied scientific endeavours, accounting for a large amount of the computational power used every day. As the power of classical computers grows, the demand in precision and scale for computational fluid dynamics increases similarly.

Future fault tolerant quantum computers promise a novel compute technology with an exponential computational power in the amount of qubits, leading to the natural questions of whether and how this novel method of computation can be used to simulate interesting problems of computational fluid dynamics (CFD). The question of a potential use for quantum computers in CFD was first researched by Yepez and his co-workers in the early 1990s, during quantum computing's first boom \cite{Yepez1998, Yepez2001, YepezBoghosian2001, Yepez2002, Pravia2003}. 
These papers describe a quantum distributed computing approach in which each grid point is described by six qubits, and the collision step at each grid point can be calculated on a separate machine. This has as a benefit that only small stable quantum computers are required and the collision step can be implemented on a quantum computer, making use of its inherently probabilistic nature. The downside of this approach is that streaming then needs to be done classically implying that complete measurement of the system and reinitialization are required in each time step. On top of that $6N$ qubits, where $N$ is the number of grid points, are required. As the number of qubits grows linearly with the size of the grid, since the grid is typically very large for CFD problems and the number of qubits currently available is very low, this poses a significant problem. 

After the initial QCFD research by Yepez et al. the field of Quantum Boltzmann methods became stagnant for over a decade. 
Whilst other QCFD approaches came to the forefront, quantum Boltzmann methods were largely forgotten until the more recent boom in 2019 starting with the paper by Todorova and Steijl \cite{Todorova2020}. 
Most recent are the methods presented in \cite{Todorova2020, Budinski2020, Budinski2021, Moawad2022, Schalkers2022, Steijl2023, Succi2023, Sanavio2023}, that all have their strengths and weaknesses. Due to the heavy computational demands of CFD and precision required, all these methods require future fault tolerant quantum computers.
The methods described in \cite{Todorova2020, Schalkers2022} include detailed quantum primitives for streaming and specular reflection but do not yet include a collision step. The methods described in \cite{Budinski2020, Budinski2021, Sanavio2023} include a quantum primitive for collision using the linear combination of unitaries approach \cite{Childs2012}, as such measurement and reinitialization are required in each time step. Due to the high computational cost of such a `stop-and-go' strategy caused by the difficulty of initialization and measurement errors, such techniques loose their practical advantage. 
The methods presented in \cite{Succi2023, Itani2023a, Sanavio2023} make use of Carleman linearization of the lattice Boltzmann equation as presented in \cite{Itani2022}. The methods of \cite{Succi2023, Itani2023a} stand out as they are geared towards quantum simulation rather than the more general quantum computation paradigm.  

What all these methods have in common is that after completing the final time step, a quantum state has been created that represents the entire flow field as a probability density distribution, e.g., encoded in the quantum state's amplitudes. So far, however, no efficient measurement strategies for this quantum state representing the flow field have been suggested. This implies that the current methods require the exponentially expensive reading out of the full quantum state to extract the entire flow field and post-process it on a classical computer afterwards. Consequently, any and all quantum advantages that were gained during the computation are lost. This paper marks the first that offers a quantum observable for the efficient reading out of the force vector acting on an object for the quantum Boltzmann method.

We first introduce the Lattice Boltzmann method in Section \ref{sec:theLBM}. In Section \ref{sec:MEM} we introduce the so-called Momentum Exchange Method (MEM) that can be used in combination with the Lattice Boltzmann method and bounce back boundary conditions to calculate the force acting on an object immersed in the fluid. Subsequently, in Section \ref{sec:QLBM} we provide the reader with the basic ideas of the Quantum Lattice Boltzmann method (QLBM) and its encoding. Using this we introduce bounce back boundary conditions for QLBM in Section \ref{sec:qbb} and ultimately in Section \ref{sec:QMEM} we introduce the Quantum Momentum Exchange Method. Finally Section \ref{sec:prac_impl} is dedicated to explaining how the QMEM can be efficiently implemented in practice and Section \ref{sec:CA} gives insight into the computational costs.

\section{The Lattice Boltzmann Method}\label{sec:theLBM}
The lattice Boltzmann method (LBM) is one of multiple widely-used computational approaches to model the behaviour of fluid flow with the aid of computers. It is based on the Boltzmann equation which can be written
\begin{equation}
  \frac{\partial f}{\partial t} + \mathbf{u}\cdot \nabla f = \Omega \left ( f \right ),
  \label{eq:boltzmann1}
\end{equation}
where $f(\mathbf{x},\mathbf{u},t)$ is the distribution function of the particle density $\rho$, over space $\mathbf{x}$, with velocity $\mathbf{u}$, at time $t$. Here, $\Omega$ represents the collision term. We furthermore assume that no external force is present.

Since the actual collision term is relatively expensive to implement, in practise typically the BGK collision term is used \cite{Bhatnagar1954}
\begin{equation}
    \Omega \left ( f \right ) = -\frac{1}{-\tau} \left ( f - f^{eq}\right ),
\end{equation}
where $\tau$ is the relaxation time and $f^{eq}$ is the equilibrium function. 
 
The Boltzmann equation can be discretized in both time, physical and velocity space leading to the lattice Boltzmann method. In the lattice Boltzmann method a time step can be denoted as
\begin{equation}
    f_i \left ( \mathbf{x} +\mathbf{c}_i \delta t, t + \delta t \right ) = f_i \left (  \mathbf{x},t \right ) - \frac{\delta t}{\tau} \left ( f_i \left (  \mathbf{x},t \right ) - f_i^{eq}\left ( \mathbf{x} ,t \right ) \right ),
\end{equation}
where subscript $i$ denotes the velocity direction.

What sets the Boltzmann method apart from other CFD approaches is that a single time step can be split into two consecutive parts, the so-called streaming and collision steps. 

Writing the state of the system after collision as $f_i^\star \left (\mathbf{x}, t \right )$ we can get
\begin{equation}
f_i^\star \left (\mathbf{x},t \right ) = f_i \left ( \mathbf{x},t \right ) - \frac{\delta t}{\tau} \left ( f_i\left (\mathbf{x},t \right ) - f_i^{eq} \left ( \mathbf{x},t \right ) \right ),   
\end{equation}
for the collision step. Subsequently the streaming step is written as
\begin{equation}
    f_i \left ( \mathbf{x}+\mathbf{c}_i\delta t, t + \delta t \right ) = f_i^\star \left ( \mathbf{x},t \right ).
\end{equation}

This ability to split the equation into two separate physically motivated steps leads to the Boltzmann method being implemented by performing collision and streaming separately and consecutively. 

\begin{figure}
     \centering
     \begin{subfigure}[b]{0.45\textwidth}
         \centering
         \includegraphics[width=\textwidth]{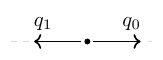}
         \caption{}
         \label{subfig:1a}
     \end{subfigure}
     \hfill
          \begin{subfigure}[b]{0.45\textwidth}
         \centering
         \includegraphics[width=\textwidth]{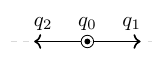}
         \caption{}
         \label{subfig:1b}
     \end{subfigure}
     \hfill
     \begin{subfigure}[b]{0.45\textwidth}
         \centering
         \includegraphics[width=\textwidth]{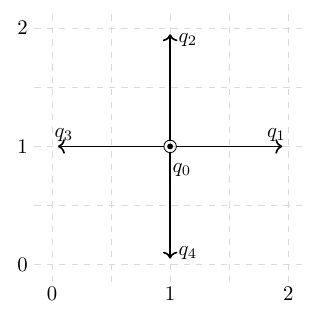}
         \caption{}
         \label{subfig:1c}
     \end{subfigure}
          \hfill
     \begin{subfigure}[b]{0.45\textwidth}
         \centering
         \includegraphics[width=\textwidth]{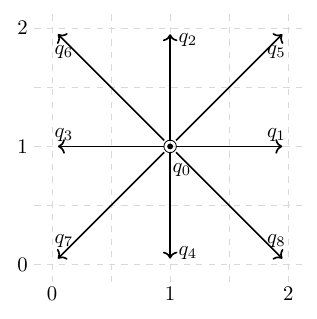}
         \caption{}
         \label{subfig:1d}
     \end{subfigure}
        \caption{Four examples of different types of D$d$Q$q$ possible. Figure \ref{subfig:1a} portrays the D1Q2 setting and Figure \ref{subfig:1b} portrays the D1Q3 setting (where a stationary particle can be included). Figure \ref{subfig:1c} portrays the D2Q5 setting and Figure \ref{subfig:1d} shows the D2Q9 setting. }
        \label{fig:multiple_DnQm}
\end{figure}

A popular way of classifying different combinations of dimensions and number of possible velocities is the so-called D$d$Q$q$ scheme. Here, $d$ denotes the number of space dimensions considered and $q$ the number of distinct velocities. In Figure \ref{fig:multiple_DnQm}
 we give four examples of different combinations of D$d$Q$q$ possible. 
In this paper we are only considering the D1Q3, D2Q9 and D3Q27 cases. 

We furthermore write $\mathbf{e}_i$ to represent the vector in the direction $i \in Q=\{0, 1, \dots,q-1\}$ of the D$d$Q$q$ scheme. For example, in the D2Q9 system we have
\begin{equation}
    \mathbf{e}_i = \begin{cases}
      (0,0) & \text{for $i = 0$}\\
      (1,0), (0,1), (-1,0), (0,-1) & \text{for $i = 1,2,3,4$}\\
      (1,1), (-1,1), (-1,-1), (1,-1) & \text{for $i=5,6,7,8$}.
    \end{cases}   
\end{equation}
Therefore $2$ qubits are necessary to represent the speed in each dimension, as the three options `positive', `negative' and `standing still' need to be encoded. 

\section{Momentum Exchange Method}\label{sec:MEM}
The momentum exchange method was proposed by Ladd \cite{Ladd1994} to determine the force acting on an object in order to calculate the drag and lift coefficients of an obstacle equipped with bounce back boundary conditions when the flow field is modelled by the Boltzmann method. Bounce back boundary conditions differ from the more intuitive specular reflection boundary conditions in that, upon contact with an object, the particle's velocity is reversed entirely instead of just the velocity normal to the object; see Figure \ref{fig:bounce_back_vs_specular_reflection}. Bounce back boundary conditions are often used in combination with the Lattice Boltzmann method \cite{Kruger2017}. In Section \ref{sec:qbb} we give an in-depth explanation of bounce back boundary conditions as well as how to implement them in our QLBM scheme.
In this paper we adopt the momentum exchange method as described in \cite{Kruger2017}. Then the force exerted on the object by the particles can be expressed as
\begin{equation}\label{eq:F1}
    \mathbf{F} = \sum_{i\in Q} \left ( \mathbf{e}_{i}f_i(\mathbf{x}_f,t) - \mathbf{e}_{\bar{i}} f_{\bar{i}}(\mathbf{x}_f,t) \right ) .
\end{equation}
In the above expression $\mathbf{x}_f$ refers to a point in the fluid space adjacent to the obstacle and $\bar{i}$ represents the velocity of the particles after particles with velocity $i$ have impinged on the object.
This expression assumes that there is no fluid inside the object and as such only takes the momentum exchange outside of the object into account. Since we are using bounce back boundary conditions we have $\mathbf{e}_{\bar{i}}:=-\mathbf{e}_i$ and
$f_i(\mathbf{x}_f,t) = f_{\bar{i}}(\mathbf{x}_f,t)$ by definition, therefore we can rewrite Equation \eqref{eq:F1} to 
\begin{equation}\label{eq:F2}
    \mathbf{F} = \sum_{i\in Q} 2 \mathbf{e}_{i} f_i(\mathbf{x}_f,t).
\end{equation}
As force is composed of magnitude and direction it is expressed by a $d$ dimensional vector with subscript $j$ denoting its $j$-th dimensional component, i.e.
\begin{equation}\label{eq:F2_d}
    F_j = \left ( \sum_{i\in Q} 2 \mathbf{e}_{i} f_i(\mathbf{x}_f,t) \right )_j.
\end{equation}

\section{Quantum Lattice Boltzmann Method}\label{sec:QLBM}
The quantum lattice Boltzmann method (QLBM) is, as the name suggests, the quantum analog of the lattice Boltzmann method. Similar to the classical lattice Boltzmann method the QLBM consists of the initialization of the problem, methods for streaming and collision, an approach to impose boundary conditions and, finally, a measurement procedure to extract application-specific QoIs. This paper introduces an efficient measurement procedure that can be used in combination with existing QLBMs. As such we abstain from presenting concrete methods for collision, streaming or state preparation. Instead we focus on explaining a set-up for the measurement procedure, which can be used with any QLBM method that uses a similar encoding scheme. 

As the measurement procedure in practice should be fitted to the quantum state that it is used on we will present how the density function is encoded in the quantum state for this method. The density function encoding presented below is similar to the ones presented in \cite{Schalkers2022, Todorova2020, Budinski2020, Budinski2021}, as such the measurement procedure presented here can be used with those papers. 

\paragraph{Flow field encoding}\label{ssec:encoding}
Building on our previous work \cite{Schalkers2022}, the quantum encoding of the discretized density function reads
\begin{equation}
    \ket{\underbrace{a_{n_a}\dots a_{1}}_{\mathrm{ancillae}} \overbrace{ g_{n_g} \dots g_{1} }^{\mathrm{position}} \underbrace{ v_{n_v} \dots v_1}_{\mathrm{velocity}}},
\end{equation}
whereby the positional and velocity qubits are split into $d$ groups, one for each dimension. More specifically, zooming in on the positional qubits, we get
\begin{equation}
\ket{g_{n_g} \dots g_1} =\ket{g^d_{n_{g_d}} \dots g^d_{1} g^{d-1}_{n_{g_{d-1}}} \dots g^{d-1}_{1} \dots  g^1_{n_{g_1}} \dots g^1_{1}},
\end{equation}
where $g^j_{n_{g_j}} \dots g^j_{1} $ encodes the $j$-th dimension of the location of grid points by representing the binary value of the location. 

Similarly if we write out the velocity qubits for the encoding explicitly we get
\begin{equation}
\ket{v_{n_v} \dots v_1} = \ket{v^d v_{\text{dir}}^d  \dots \dots v^1 v_{\text{dir}}^1},
\end{equation}
where $v_{\text{dir}}^j$ expresses the direction (positive or negative) of the particle in dimension $j$ and the $v^j$ qubits express whether a particle has a nonzero velocity in dimension $j$. Notice that this order is different from the one presented in \cite{Schalkers2022} where the $v_\text{dir}$ qubits are grouped together, this is done simply to make the observable in Section \ref{subseq:obs} easier to visualize as a matrix. Another difference from the setup presented in \cite{Schalkers2022} is that here we are only considering the D1Q3, D2Q9 and D3Q27 cases leading to exactly two velocity qubits per dimension.

The ancilla qubits are used for several different purposes throughout the QLBM method. In this paper we will only highlight the labels and purposes of the ancillae that are used in the quantum bounce back boundary conditions implementation and the quantum momentum exchange method presented in Sections \ref{sec:qbb} and \ref{sec:QMEM}, respectively. 
We identify the $a_{v,i}$ ancilla qubits that indicate whether in this time step the associated particles are streamed in dimension $i$. Furthermore we make use of the $a_{o}$ which is the ancilla qubit that indicates whether or not a particle is in an object and the bounce back boundary conditions need to be applied. 

\section{Quantum bounce back boundary conditions}\label{sec:qbb}
\begin{figure}
    \centering
\includegraphics[]{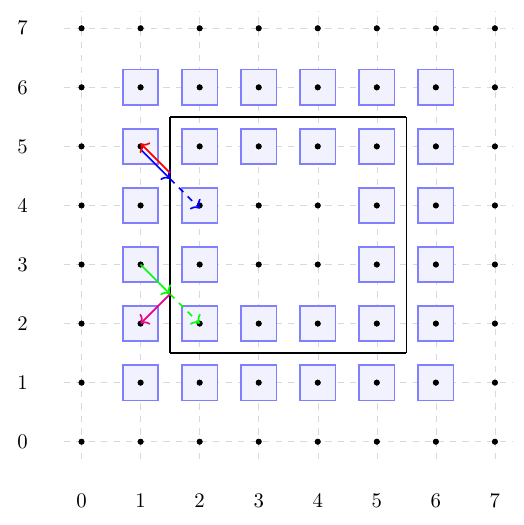}
    \caption{Illustration of bounce back boundary conditions (top, in red and blue arrows) versus specular reflection boundary conditions (bottom, in green and magenta).}
    \label{fig:bounce_back_vs_specular_reflection}
\end{figure}

One of the most commonly used boundary conditions in practical LBM is the bounce back boundary condition which amounts to fully reflecting the direction of particles that get into contact with obstacles and resetting them to their original position inside the flow domain \cite{Schiller2008, Kruger2017}. This is different from the specular reflection boundary conditions that we adopted in our earlier work \cite{Schalkers2022}, which reverses only the normal component of the velocity vector. Figure \ref{fig:bounce_back_vs_specular_reflection} illustrates the difference between the two types of boundary conditions. 

The algorithm for implementing bounce back boundary conditions in a classical LBM can be summarized as follows. First, the particles that virtually travelled into the obstacle have their velocity direction reversed in all dimensions and subsequently these particles are placed outside of the obstacle. The particles are placed in the correct position outside of the obstacle by moving one grid point in the dimension(s) that they previously moved in. 

\begin{figure}
    \hspace{-4cm}
    \includegraphics[]{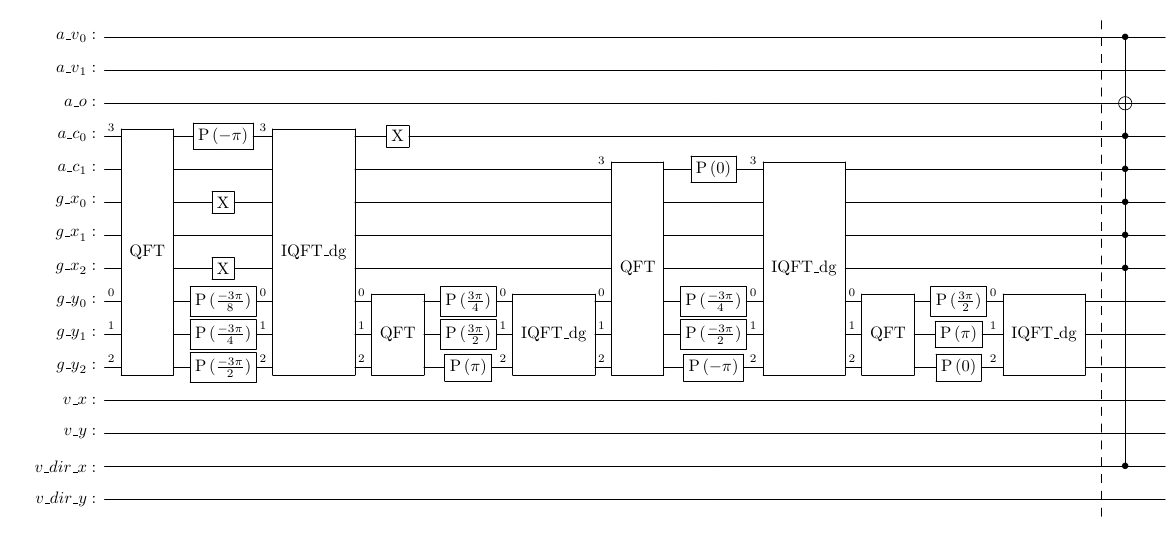}
    \caption{The first part of the bounce back boundary conditions, applied to the example of Figure \ref{fig:example_right} to properly reset the particles moving to the right in the x-direction hitting the particles on the left wall. This part of the algorithm sets the ancilla qubit, indicating that particles have virtually travelled into the object and need to have their velocity reversed and be moved out.}
    \label{fig:bb-wall-refl}
\end{figure}

\begin{figure}
    \centering
    \includegraphics[]{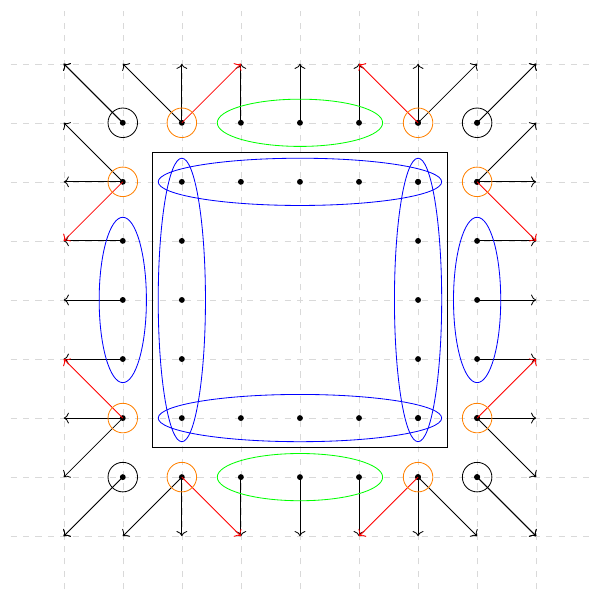}
    \caption{Illustration of all possible corner cases to be taken into account when particles collide with an obstacle (black box). The arrows and colors indicate the physically correct behavior for a fail-safe implementation of the boundary conditions.}
    \label{fig:example_right}
\end{figure}

For implementing the bounce back boundary condition as a quantum primitive we require only one ancilla qubit $a_{o}$ to indicate whether a particle has virtually moved into an object. The ancilla $a_{o}$ is initialized in the $\ket{0}$ state and flipped to $\ket{1}$ when a particle has virtually travelled into one of the points inside the obstacle. We check whether a particle has virtually travelled into the object using the efficient object encoding method as described in Section 5.4 of \cite{Schalkers2022}. In Figure \ref{fig:bb-wall-refl} we show how this efficient object encoding method can be implemented for the example give in \ref{fig:example_right}. In Figure \ref{fig:bb-wall-refl} we show the quantum comparison operation that can check whether or not a particle has come into contact with the wall from (2,2) to (2,5). This is done by checking whether the location on the x-axis is equal to two as is done by applying an X gate to the $g_{x_0}$ and $g_{x_2}$ qubit. We use two quantum comparison operations to check whether $2 \leq y \leq 5$ as can be seen in the picture by the QFT operations followed by rotations and IQFT. The mathematics behind this procedure is explained in Section 5.4 of \cite{Schalkers2022}.

\begin{figure}
    \hspace{-4cm}
    \includegraphics[]{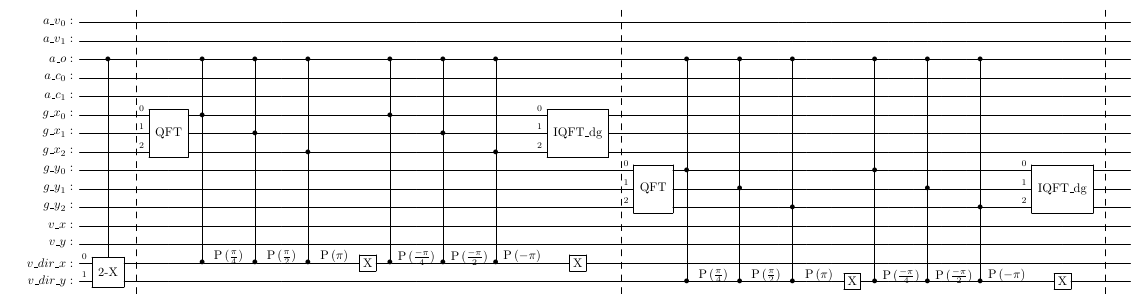}
    \caption{This figure shows part of the bounceback boundary conditions quantum algorithm where the velocity qubits of the particles hitting the object get reversed and subsequently moved back out of hte object.}
    \label{fig:bb-refl-and-stream}
\end{figure}

\begin{figure}
    \hspace{-4cm}
    \includegraphics[]{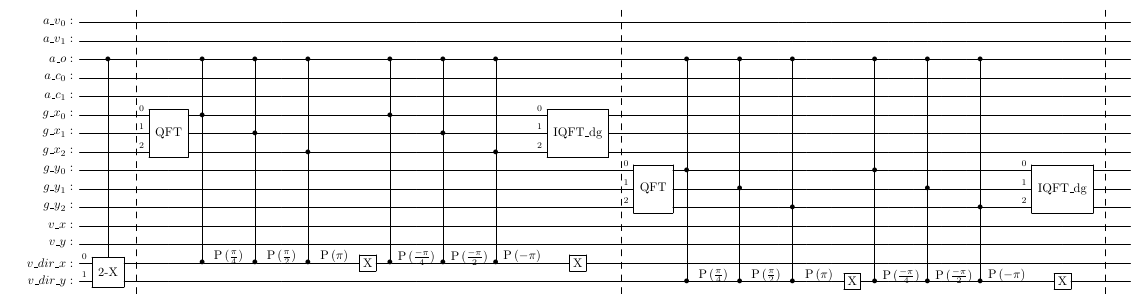}
    \caption{This figure shows part of the bounceback boundary conditions quantum algorithm where the ancilla qubit $a_O$ qubits of the particles that hit the object and got reversed and subsequently moved back out of hte object gets reset.}
    \label{fig:bb-cornerpoints}
\end{figure}

As a next step we flip the state of the $v_\text{dir}^j$ qubits for all dimensions $j$ controlled on the state of the $a_o$ ancilla. By doing this we make sure that the velocity direction is reversed in all dimensions after contact with an obstacle as is required for bounce back boundary conditions. And subsequently the particles are moved by one position controlled on the $a_v^j$, $v_\text{dir}^j$, and $a_o$ qubits to ensure that the particles move one step in the correct direction in the dimension(s) that they moved in when they moved into the obstacle and of course to ensure that this only happens after the particles moved into the obstacle.

This is done by the operation controlled double NOT operation shown in the beginning of Figure \ref{fig:bb-refl-and-stream}. Subsequently, as can be seen in the same figure, controlled or whether or not the $a_{o}$ qubit is in the state 1, we stream in the x and y dimension, thereby making sure we only stream the particles that just collided with the object. This is done to ensure that the particles are set back outside of the object again.

Finally the $a_o$ qubits need to be reset to $\ket{0}$ before we can start the next time step. As in our previous work \cite{Schalkers2022} the blue and green encircled points outside of the object constitute the trivial case in Figure \ref{fig:example_right}. We reset the $a_o$ qubits controlled on if we are in one of the blue (green) encircled points, the direction of the $x$ ($y$) velocity and the ancilla qubit indicating whether we moved in the dimension in this time step $a_v^1$ ($a_v^2$). Specifically we reset the ancilla qubit $a_o$ if we are in a blue (green) encircled grid point outside the object and $a_v^1 = 1$ ($a_v^2 = 1$) and $v_\text{dir}^1$ ($v_\text{dir}^2$) points away from the object. Using this logic the $a_o$ qubits are reset to $\ket{0}$ for each wall separately.

Resetting the $a_o$ qubit is a bit more difficult around the edges as indicated with black and orange encircled points in Figure \ref{fig:example_right}. 

For the black encircled points outside the corner we need to reset the ancilla qubit $a_o$ if and only if both $v_\text{dir}^1$ and $v_\text{dir}^2$ point in the direction away from the object and $a_v^1 = a_v^2 =1$ holds.

As for the orange encircled `side-edge' grid point we first reset the $a_o$ ancilla in the same way as for the blue (green) encircled points described above. Now we only need to note that for the case described by the red arrow in Figure \ref{fig:example_right} we have wrongly flipped the $a_o$ ancilla qubit and so we need to verify whether we are in the `red arrow' case by checking if $v_\text{dir}^1$ and $v_\text{dir}^2$ pointed in the direction of the arrow and if $a_v^1 = a_v^2 =1$ holds. If the particle is in a state where  $a_v^1 = a_v^2 =1$ and $v_\text{dir}^1$ and $v_\text{dir}^2$ are such that the particle is in an red arrow case the $a_o$ ancilla get flipped again, back to the original state of $\ket{0}$. 

In Figure \ref{fig:bb-cornerpoints} we show how the ancilla qubits are reset. In this picture it can be seen that controlled on the conditions described above, an X gate is applied to the ancilla qubits to re-set them to the zero state if and only if they were in the one state to begin with. Resetting the ancilla qubits in this particular use case is non-trivial as the ancilla qubits were originally set based on location and then streamed, so we need to keep track on how all the particles moved in order to be able to reset them properly. The core is to now check if the particles are just outside the object pointing away from the object in such a way that could only have happened if they just reflected from the object. We can safely reset the ancilla qubits based on this, the circuit is given in figure \ref{fig:bb-cornerpoints}. 

\section{Quantum momentum exchange method}\label{sec:QMEM}
In this section we explain how the momentum exchange method can be expressed as an observable for the encoding described in Section \ref{ssec:encoding}. In order to do this we will first change the density encoding of the quantum state into a rooted density encoding. Using this rooted density encoding we can subsequently define the observable that calculates the force using Equation \eqref{eq:F2} and finally we describe how this method can be implemented as an executable quantum circuit.

\paragraph{Rooted density encoding}\label{subseq:rooted}
Since the momentum exchange method is linear in nature, whereas quantum observables are quadratic, it is advisable to change from encoding the density function $ f_i \left  (\mathbf{x},t \right ) $ in the quantum state $\ket{\psi}$ to an encoding of the square root of the density function $\sqrt{ f_i \left ( \mathbf{x}, t \right ) }$. Such a shift to a rooted density encoding can be done without altering any subsequent circuits in the methods \cite{Schalkers2022, Todorova2020}. In practice the densities should be rooted before the start of the quantum algorithm, therefore the information will be initialized into the quantum state such that the density is already rooted and no quantum method is required for this task.

\subsection{Momentum exchange method as an observable}\label{subseq:obs}
We now derive the observable that calculates the force exerted on the object using the momentum exchange method. As force is described using a vector, we need to calculate the values of the vector for all $d$ dimensions. Here we show how to calculate this vector using $2d$ different observables, where each observable calculates the value of the force vector in one spatial dimension, $d\in\{1,2,3\}$, and one velocity direction. This is done to make the resulting observable easier to portray and explain, since all the observables can be expressed as diagonal matrices they do commute and so they can all be measured in the same runs. 

Considering the encoding described above, equation \eqref{eq:F2} can be evaluated from the quantum state $\ket{\psi}$ encoding $\sqrt{ f_i \left (\mathbf{x},t \right )}$, by the diagonal observable $O_{\text{OME}}$ to be specified below. The diagonal of the matrix expressing the observable $O_\text{OME}$ is built up using $B_{\text{OME}}$ matrices, which are placed at grid points in the fluid domain directly adjacent to the obstacle. All the other indices of the matrix matrix expressing $O_{\text{OME}}$ will remain zero. An example of this is the matrix
\begin{equation}\label{eq:Obs_1}
        O_{\text{OME}} = \begin{bmatrix}
            \dots & \dots & \dots & \dots  \\
            \dots &  B_{\text{OME}} & \dots & \dots \\
            \dots & \dots & \dots & \dots \\
            \dots & \dots & \dots & \dots \\
        \end{bmatrix},
\end{equation}
where the dots represent $4 \times 4$ matrices with only 0 indices and $B_{\text{OME}}$ is a $4 \times 4$ matrix that can be written as 
\begin{equation}\label{eq:bmatrix}
    B_{\text{OME}} = \begin{bmatrix}
        0 & 0 & 0 & 0\\
        0 & 0 & 0 & 0 \\
        0 & 0 & 2 & 0 \\
        0 & 0 & 0 & 0 
    \end{bmatrix}.
\end{equation}

\begin{figure}
    \centering
    \includegraphics[]{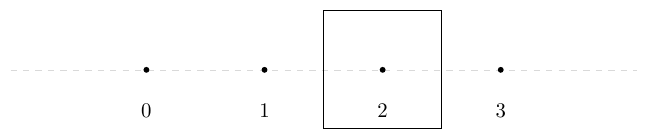}
\caption{Example of the D1Q3 case with four grid points and one obstacle located on the third grid point.}
\label{fig:D1Q3ex_obs_4g_1b}
\end{figure}

The example given in Equation \eqref{eq:Obs_1} represents a 1-dimensional case with 4 grid points and three possible speeds (one in the positive and one in the negative $x$-direction as well as the zero speed) and the wall adjacent to the second grid point as represented in Figure \ref{fig:D1Q3ex_obs_4g_1b}. Since $B_{\text{OME}} = B_{\text{OME}}^{\dagger}$, both $B_{\text{OME}}$ and $O_{\text{OME}}$ are Hermitian and therefore $O_{\text{OME}}$ constitutes a quantum observable. It can easily be seen that any other distribution of $B_{\text{OME}}$ along the diagonal will also lead to a quantum observable.

For the D1Q3 case with four grid points described above we have the following quantum state encoding \begin{equation} \ket{\psi} = \frac{1}{\sqrt{\sum_{x,i} f_i \left (\mathbf{x},t \right )}} \sum_{x,i} \sqrt{ f_i \left (\mathbf{x} ,t \right )}\ket{g^1_2g^1_1v^1v^1_\text{dir}},
\end{equation} where $g^1_2g^1_1$ represent the binary value of the location of the $x$-axis, $\ket{v^1v^1_\text{dir}} = 10$ indicates streaming in the positive $x$-direction, $\ket{v^1v^1_\text{dir}}=11$ indicates streaming in the negative $x$-direction and $\ket{v^1v^1_\text{dir}}=00$ as well as $\ket{v^1v^1_\text{dir}}=01$ indicates that the particle is not streaming in the $x$-direction. Here and in the remainder of this Section we do not take into account the ancilla qubits as they play no role in the final density function and as such will not be part of the measurement process.

With the above convention, the quantum state can be written as the coefficient vector relative to the computational basis as follows: 
\begin{equation}\label{eq:psi}
    \ket{\psi} = \sum_{x,v} \alpha_{x,v} \ket{g^1_2g^1_1v^1v^1_\text{dir}} = \frac{1}{\sqrt{\sum_{x,i} f_i \left (\mathbf{x},t \right )}} \begin{bmatrix}
        0 \\
        \sqrt{f_0(0,t)} \\
        \sqrt{f_1(0,t)} \\
        \sqrt{f_2(0,t)} \\
        0 \\
        \sqrt{f_0(1,t)} \\
        \sqrt{f_1(1,t)} \\
        \sqrt{f_2(1,t)} \\
        0 \\
        \sqrt{f_0(2,t)} \\
        \sqrt{f_1(2,t)} \\
        \sqrt{f_2(2,t)} \\
        0 \\
        \sqrt{f_0(3,t)}  \\ 
        \sqrt{f_1(3,t)} \\
        \sqrt{f_2(3,t)}         
    \end{bmatrix}.
\end{equation}
Using expression \eqref{eq:psi} and some basic linear algebra it follows that
\begin{equation}\label{eq:obs_value}
    \braket{\psi|O_{\text{OME}}|\psi} = \frac{2f_1 \left ( 1, t \right )}{\sum_{x,i}f_i \left (\mathbf{x},t \right )} .
\end{equation}
Since the value of $\sum_{x,i}f_i \left (\mathbf{x},t \right )$ is known when starting the algorithm we can simply multiply Equation \eqref{eq:obs_value} by $\sum_{x,i}f_i \left (\mathbf{x},t \right )$ to find the value of the force we wish to calculate as described in Equation \eqref{eq:F2}, which can subsequently be used to calculate the drag and lift coefficient \cite{Ladd1994}. 

\paragraph{Extension to more dimensions}
This method can easily be extended to more dimensions by noticing that the $B_\text{OME}$ matrix is of size $2^{n_v} \times 2^{n_v}$ and should consist of only one non-zero element. This non-zero element will always be placed on the diagonal at the position of the basis state $\ket{v^i}$. 

\paragraph{Complexity Analysis}\label{sec:CA}
The number of measurements required to determine the force with an $\epsilon$ precision using our proposed approaches depends on multiple factors. 
As long as the total number of grid points located inside and adjacent to the boundary of the obstacle is polynomial in the total number of grid points, the number of diagonal elements that are non-zero in the observable is polynomial in the size of the grid. This means that the number of non-zero elements in the observable is not exponentially small in the total size of the system. Therefore, in this case, we wish to measure a subspace that is only polynomially small in the total size of the system which is feasible without exponential overhead. 

\section{Practical implementation of the momentum exchange method on a quantum computer}\label{sec:prac_impl}
Realizing an observable on a real-world quantum computer amounts to implementing a quantum circuit that translates the observable to measurements in the computational Z-basis.
We will now present the quantum circuit that translates the observable described in Subsection \ref{subseq:obs} for determining the force in one dimension in one direction to measuring one qubit in the Z-basis, making the process clear and easily implementable on a quantum device. 

We will first describe how this operation can be applied by using an already implemented circuit for the bounce back boundary condition and we will subsequently show that this operation indeed transforms the described observable to one that consists of a Z measurement on one qubit. 

\subsection{Implementation using ancilla qubits for bounce back boundary conditions}
We have implemented a method to measure the expectation value of the described observable by measuring only one qubit. This is done using the implementation of the bounce back boundary conditions. In this implementation a qubit $a_{o}$ gets flipped to indicate that a particle is inside an object. To determine the expectation value of the observable we will use these ancilla qubits differently. We will from now on call this $a_{o}$ ancilla qubit that was used for the bounce back boundary conditions $a_{o,+}$ and we define a second ancilla qubit $a_{o,-}$. These ancilla qubits will be flipped if a force was exerted on the object in a positive or negative direction, respectively. In order to do this we apply a multi-controlled NOT operation controlled on the  qubits to determine whether we are in the object and the qubit indicating the direction of the particles in the dimension to the $a_{o,+}$ ($a_{o,-}$) qubits.  

\begin{figure}
    \hspace{-4cm}
    \includegraphics[]{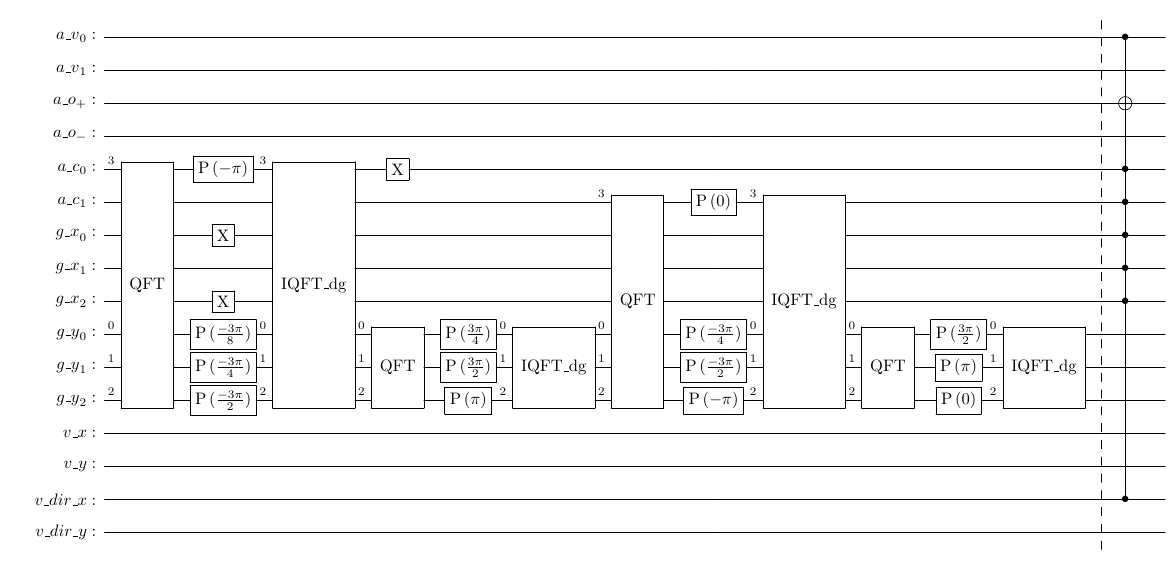}
    \caption{The quantum algorithm for the QMEM, applied to the example of Figure \ref{fig:example_right} to determine the force of the particles moving to the right in the x-direction hitting the particles on the left wall. This part of the algorithm sets the ancilla qubit, indicating that particles have virtually travelled into the object. Subsequently the ancilla qubit $a_{o, +}$ is measured to determine the force.}
    \label{fig:qmem_qcirc}
\end{figure}

By doing this we are extracting the relative density of particles that come into contact with an obstacle in the positive and negative direction for the considered dimension. Subsequently we measure the qubits $a_{o,+}$ and $a_{o,-}$ and subtract the expectation value of $a_{o,-}=1$ from $a_{o,+} = 1$. The resulting value expresses the relative pressure in the positive / negative direction. Figure \ref{fig:qmem_qcirc} shows the quantum circuit that implements this quantum momentum exchange method for the example of Figure \ref{fig:example_right}.

\subsection{Proof of method}
In this section we show that using the method described above, we indeed calculate the force exerted on an object in one dimension as expressed in Equation \eqref{eq:F2}.

We flip the $a_{o,+}$ ancilla qubit in the case that particles have impinged on the object and the particles have a positive velocity in the $x$-direction. 
Therefore, after re-arranging some qubits, we can write 
\begin{equation}
\begin{split}
    & \bigl( \sqrt{f_{v_0} \left ( x_0,t \right )} \ket{ \left ( g_{n_g} \dots g_{1} \right )_0 \left (  v_{n_v} \dots v_1 \right )_0 } + \dots +   \\ & \sqrt{f_{v_1} \left ( x_1, t\right )} \ket{ \left (  g_{n_g} \dots g_{1} \right )_1 \left (  v_{n_v} \dots v_1 \right )_1} \bigr) \ket{0}_{a_{o,+}} +  \\ & \bigl( \sqrt{f_{v_2} \left ( x_2, t \right )}\ket{\left ( g_{n_g} \dots g_{1} \right )_2 \left ( v_{n_v} \dots v_1 \right)_2} + \dots +  \\  & \sqrt{f_{v_3} \left ( x_3 ,t\right )} \ket{ \left ( g_{n_g} \dots g_{1} \right )_3 \left ( v_{n_v} \dots v_1 \right )_3}  \bigr) \ket{1}_{a_{o,+}},
    \end{split}
\end{equation}
where for $v_i$ and $x_i$  the subscript is simply used to indicate that a specific value for the location and velocity is considered and similarly for the qubits $\left ( g_{n_g} \dots g_{1} \right )_i$ $\left (  v_{n_v} \dots v_1 \right )_i$ the subscript is used to indicate the velocity and grid point qubits are in a specific state. 
Here the states $ \ket{ \left ( g_{n_g} \dots g_{1} \right )_0 \left (  v_{n_v} \dots v_1 \right )_0 } + \dots +  \ket{\left (  g_{n_g} \dots g_{1} \right )_1 \left (  v_{n_v} \dots v_1 \right )_1}$ describe exactly the particles that do not impinge on the object in the positive $x$-direction in the current time step, as the $a_{o,+}$ qubit is in the $\ket{0}$ state. Similarly the states $\ket{\left ( g_{n_g} \dots g_{1} \right )_2 \left ( v_{n_v} \dots v_1 \right)_2} + \dots + \ket{\left ( g_{n_g} \dots g_{1} \right )_3 \left ( v_{n_v} \dots v_1 \right )_3}$ represent the relative densities of the particles that have impinged the object in the positive $x$-direction.
From this we can conclude that the total probability of finding $\ket{a_{o,+}} = \ket{1}$ upon measurement is equal to \begin{equation}
    f_{v_2} \left ( x_2, ,t\right ) + \dots + f_{v_3} \left ( x_3,t \right ),
\end{equation}
which is precisely equal to the relative density of particles hitting the object with a positive velocity and which is precisely what we wish to measure. 

\section{Conclusion}
In this paper we have presented a quantum approach to determine the force of the flow field acting on an object immersed in the fluid via an efficient and easily implementable measurement procedure for the quantum lattice Boltzmann method. To the best of our knowledge, this is the first time that efficient measurement strategies are addressed in the QLBM literature. Previous works are limited to reading out the entire flow field which cannot be realized efficiently on a quantum computer, thereby destroying any quantum advantage.

Our approach represents the quantum analog of the momentum exchange method and consists of a quantum primitive for implementing bounce back boundary conditions at the end of each time step and an observable that can be easily implemented as measurements in the computational basis to obtain the forces exerted by the fluid on an internal object.

\section{Acknowledgement}
We gratefully acknowledge support from the joint research program “Quantum Computational Fluid Dynamics” by Fujitsu Limited and Delft University of Technology, co-funded by the Netherlands Enterprise Agency under project number PPS23-3-03596728.
Furthermore the authors would like to thank C\u{a}lin Georgescu for his feedback on the manuscript. 

\printbibliography

\end{document}